\documentclass[aps,pra,superscriptaddress,amsfonts,amssymb,amsmath,eqsecnum,nofootinbib,twocolumn,floatfix]{revtex4-2}
\usepackage{color,graphicx}
\usepackage[utf8]{inputenc}
\usepackage[T1]{fontenc}
\usepackage[english]{babel}
\definecolor{darkblue}{rgb}{0,0,0.7}
\definecolor{darkred}{rgb}{0.7,0,0}
\definecolor{darkgreen}{rgb}{0,0.4,0}
\usepackage[unicode, colorlinks, citecolor=darkblue, linkcolor=darkred, urlcolor=blue]{hyperref}

\allowdisplaybreaks
\begin{document}
	
	\title{Noiseless signal amplification in an opto-mechanical transducer}
	
	\author{Aleksandr A. Movsisian}
	
	\affiliation{Faculty of Physics, M.V. Lomonosov Moscow State University, Leninskie Gory, Moscow 119991, Russia}
	
	\author{Albert I. Nazmiev}
	
	\affiliation{Faculty of Physics, M.V. Lomonosov Moscow State University, Leninskie Gory, Moscow 119991, Russia}
	
	\author{Andrey B. Matsko}
	
	\affiliation{Jet Propulsion Laboratory, California Institute of Technology, 4800 Oak Grove Drive, Pasadena, California 91109-8099, USA}
	
	\author{Sergey P. Vyatchanin}
	\affiliation{Faculty of Physics, M.V. Lomonosov Moscow State University, Leninskie Gory, Moscow 119991, Russia}
	\affiliation{Quantum Technology Centre, M.V. Lomonosov Moscow State University, Leninskie Gory, Moscow 119991, Russia}
	\affiliation{Faculty of Physics, Branch of M.V. Lomonosov Moscow State University in Baku,\\ 1 Universitet street, Baku, AZ1144, Azerbaijan}

\begin{abstract}
The high-sensitivity quantum detection of a resonant classical force acting on a quantum oscillator can be substantially enhanced through the use of a resonant optical parametric transducer. We demonstrate that this approach not only enables quantum back-action evasion measurements that exceed the Standard Quantum Limit of sensitivity but also facilitates the noiseless amplification of the classical signal. This amplification is achieved by independently measuring the two modulation sidebands generated by the signal force, allowing for a more precise and noise-resistant detection process.
\end{abstract}



\maketitle
\section{Introduction}

The sensitivity of the detection of a classical mechanical force is usually constrained by quantum optical noise, setting the well-known Standard Quantum Limit (SQL) \cite{Braginsky68,92BookBrKh}. The SQL arises from the non-commutativity of the optical probe noise and the quantum back-action noise, which stems from the ponderomotive effect of the probe light on the mechanical system. The optimal measurement approaches that allow surpassing of the SQL fall into two distinct subclasses based on the type of signal force. The free-mass mechanical probe configuration is used for detecting broadband mechanical forces, while the resonant mechanical configuration is suited for detecting narrowband signal forces. Here we study the second case and consider the detection of two modulation optical sidebands that escape the mechanical system, noting that a dichromatic technique can lead to the observation of phenomena such as negative radiation pressure \cite{Povinelli05ol,maslov13pra} and evasion of the optical quadrature-dependent quantum back action \cite{21a1VyNaMaPRA}.  

\begin{figure}[ht]
\includegraphics[width=0.45\textwidth]{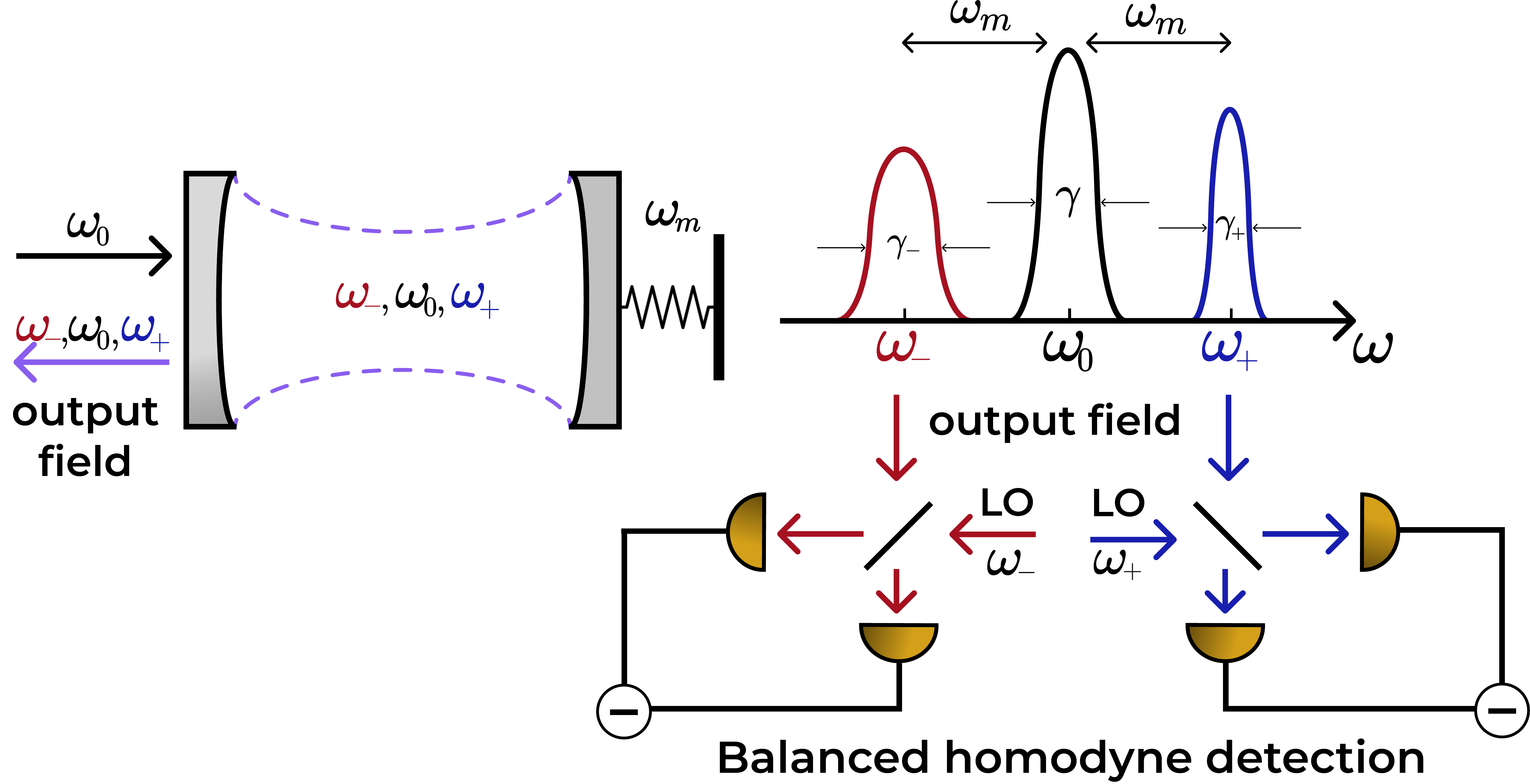}
\caption{Frequencies $\omega_0,\ \omega_\pm$ of three optical modes in the FP cavity are separated by the frequency $\omega_m$ of mechanical oscillator. Optical modes are coupled to the mechanical oscillator via ponderomotive pressure. The relaxation rates of the optical modes are smaller than mechanical frequency $\gamma,\ \gamma_\pm\ll \omega_m$. The central mode with frequency $\omega_0$ is resonantly pumped. The outputs of modes $\omega_\pm$ are detected separately.}\label{schemeM}
\end{figure}

We investigate the optimization and practical implementations of quantum measurements the nearly resonant force of interest acts on a mechanical oscillator represented by a mirror of a Fabry-Perot (FP) cavity. The motion of the mirror is tracked using three optical modes of the cavity. The frequencies of the optical modes $\omega_\pm,\ \omega_0$ are separated by the mechanical frequency $\omega_m$ (Fig.~\ref{schemeM}), so the equities $\omega_\pm =\omega_0\pm \omega_m$ remain valid. The measurement is performed by optical pumping of the central optical mode $\omega_0$ and measuring the light that escapes two other modes $\omega_\pm$. Detection of optimal quadrature components of the output waves of modes $\omega_\pm$ {\em separately} provides two registration channels. The signal is obtained through post-processing by taking a weighted linear combination of the measured results. It allows detecting back action and to remove it {\em completely in a broad band} from the measured data via post-processing if the modes have the same bandwidth and identical geometry \cite{21a1VyNaMaPRA}. 

Here we have identified a unique combination of optical relaxation rates for modes $\omega_\pm$ and opto-mechanical coupling coefficients that allows for the complete elimination of quantum back-action effects from the measurement results. Furthermore, beyond mitigating quantum back-action, our approach also {\em removes any influence of optical shot noise on the measurement outcomes}. Additionally, the measured signal undergoes amplification, further enhancing the technique's effectiveness. These three key advantages—elimination of quantum back-action, suppression of optical shot noise, and signal amplification—demonstrate that our method surpasses traditional QND measurement techniques in efficiency. To our knowledge, such a measurement strategy was not previously studied.

\section{Model and results}

We assume that the relaxation rates of the optical modes (Fig.~\ref{schemeM}) are different and characterized by the full width at the half maxima (FWHM) defined by  $2\gamma_\pm =2(\gamma_{0\pm}+\gamma_{e\pm})$ and $
2\gamma = 2(\gamma_0+\gamma_e)$, for the modes $\omega_\pm$ and for $\omega_0$, respectively. Here $\gamma_{0\pm},\ \gamma_0$ characterize the effective transmittance of the input mirror and $\gamma_e,\ \gamma_{e\pm}$ stand for optical losses of the cavity. We assume that the relaxation rates $\gamma_e,\ \gamma_{e\pm}$ are small compared to transmission ones: $ \gamma_e,\ \gamma_{e\pm} \ll \gamma_0,\ \gamma_{0\pm}$. The mechanical relaxation rate, $\gamma_m$, is small compared to the optical relaxation rate. We also assume that the optical triplet is symmetric and tuned to match the mechanical oscillator. Additionally, the conditions for resolved sideband interaction and relatively small optical loss are satisfied. $\omega_\pm =\omega_0\pm \omega_m$, and $\gamma_m\ll \gamma,\  \gamma_\pm \ll \omega_m$. 

The generalized Hamiltonian describing the system can be presented in form
\begin{subequations}
   \label{Halt}
  \begin{align}
  H   &= H_0 + H_\text{int}+H_s + \\
  & +H_{T 0}+H_{\gamma_0} + H_{T e}+H_{\gamma_e}+H_{T, \, m}+H_{\gamma_ m} ,\nonumber\\
  \label{H0}
  H_0 &=\hslash \omega_+\hat c_+^\dag \hat c_+  + \hslash \omega_0 \hat c_0^\dag \hat c_0  + \hslash \omega_-\hat c_-^\dag \hat c_-
         +\hslash \omega_m \hat d^\dag \hat d,\\
  \label{Hint}
    H_\text{int} & = \frac{\hslash }{i}
      \left( \left[\eta_-\hat c_0^\dag \hat c_- + \eta_+\hat c_+^\dag \hat c_0\right] \hat d -  adj. \right),\\
   \label{Hs}\hat 
   H_s & = - F_s x_0\left(\hat d+ \hat d^\dag\right). 
  \end{align}
  \end{subequations}
Here, $\hslash$ is the Planck constant. The term $H_0$ represents the energies of the optical modes and the mechanical oscillator. The operators $\hat c_0, \ \hat c^\dag_0, \ \hat c_\pm, \ \hat c^\dag_\pm$ are the annihilation and creation operators for the corresponding optical modes, while $\hat d$ and $\hat d^\dag$ are the annihilation and creation operators for the mechanical oscillator. The coordinate operator $\hat x$ of the mechanical oscillator is expressed in the form $\hat x= x_0\left(\hat d + \hat d^\dag\right)$, where $x_0=(\hslash/(2m \omega_m))^{1/2}$, $m$ is the mass of the oscillator. The term $H_\text{int}$ is the interaction Hamiltonian, written in the rotational wave approximation for the optical and mechanical modes. It is worth noting that the canonical interaction Hamiltonian, $H_\text{int} \sim (E_0 + E_+ + E_-)^2 x$, where $E_0, E_-, E_+$ are the electric fields of modes $0, -, +$ at the mirror surface (the oscillator mass), can be transformed into the form \eqref{Hint} after neglecting rapidly oscillating terms. We consider here a non-symmetric interaction by introducing two distinct coupling constants, $\eta_\pm \simeq x_0 \omega_0 / L$, where $L$ is the length of the cavity. The term $H_s$ represents the Hamiltonian for the signal force $F_s$. The Hamiltonian $H_{T0}$ describes the external environment (including regular and fluctuating fields incident on the input mirror), while $H_{\gamma_0}$ accounts for the coupling between the external environment and the optical modes, resulting in a decay rate $\gamma_0$; $H_{Te}$ and $H_{\gamma_e}$ represent optical losses. The pump is also incorporated into $H_{\gamma_0}$. Similarly, $H_{T, , m}$ is the thermal bath Hamiltonian, and $H_{\gamma_m}$ describes the coupling between the environment and the mechanical oscillator.

In what follows we analyze the opto-mechanical system and find that the back action can be eliminated if 
\begin{equation} \label{zerodamp}
    g_+=g_-,\quad g_\pm =C_0^2\frac{\eta_\pm^2}{\gamma_\pm},
\end{equation}
$|C_0|^2 \gg 1$ is an expectation value of the number of photons in the optically pumped mode. The condition (\ref{zerodamp}) corresponds to the case of complete suppression of the mechanical damping impinged on the movable mirror due to asymmetry of the system. 

This is not the optimal selection of the coefficients, though. The best fundamentally achievable sensitivity of the measurement is reached when the damping coefficient of the mechanical mode coincides with nonlinear mechanical attenuation $\gamma_m=G$, where $G=g_+-g_-$
At the same time the signal experiences unconstrained amplification proportional to the photonic gain coefficient $G_+=g_++g_-$.

To perform the measurements, we need to analyze the light exiting the FP cavity. To determine these fields, we employ the input/output formalism using the Langevin approach. We denote the input and output optical amplitudes as $\hat a_{\pm, ,0}$ and $\hat b_{\pm, ,0}$, respectively. Starting from the Hamiltonian \eqref{Halt}, we derive the equations of motion for the intracavity slow amplitudes of the fields.
\begin{subequations}
 \label{moveq}
\begin{align} 
	\label{c0}
\dot {\hat c}_0+\gamma \hat c_0&=\eta_+^*\hat c_+ \hat d^\dag - \eta_- \hat c_- \hat d+  +\sqrt{2 \gamma_0}\,\hat a_0 + \sqrt{2 \gamma_e}\, \hat e_0,\\
\dot {\hat c}_++\gamma_+ \hat c_+&=-\eta_+ \hat c_0 \hat d 
	+\sqrt{2 \gamma_{0+}}\,\hat a_+ +\sqrt{2\gamma_{e+}} \hat e_+, \\
\dot { \hat c}_-+\gamma_- \hat c_-&=\eta_-^*\hat c_0 \hat d^\dag 
	 + \sqrt{2 \gamma_{0-}}\,\hat a_-+\sqrt{2 \gamma_{e-}}\,\hat e_-,\nonumber\\
\dot {\hat d}+\gamma_m \hat d&=\eta^*_-\hat c_0 \hat c_-^\dag + \eta^*_+\hat c_0^\dag \hat c_+  +\sqrt{2 \gamma_m}\,\hat q +f_s.
\end{align} 
\end{subequations} 
Here, the operators $\hat e_\pm$ represent quantum fluctuations due to optical losses, $\hat q$ is the fluctuation force acting on the mechanical oscillator, and $f_s$ is the signal force.  

The operators $\hat a_\pm,\ \hat e_\pm,\ \hat q$ are characterized by the standard commutators and correlators, $\left[\hat a_\pm(t), \hat a_\pm^\dag(t')\right] = \left\langle\hat a_\pm(t)\, \hat a_\pm^\dag(t')\right\rangle = \delta(t-t')$, $\left[\hat q(t), \hat q^\dag(t')\right] =   \delta(t-t')$, and $\left\langle\hat q(t) \hat q^\dag(t')\right\rangle = (2n_T +1)\, \delta(t-t')$.
Here, $\langle \dots \rangle$ denotes ensemble averaging. The term $n_T$ represents the thermal number of mechanical quanta.

The input-output relations that connect the incident ($\hat a_\pm$), intracavity ($\hat c_\pm$), and output ($\hat b_\pm$) amplitudes are given by
\begin{align}
 \label{outputT}
  \hat b_\pm= -\hat a_\pm + \sqrt{2\gamma_{0\pm}} \hat c_\pm.
\end{align} 
It is convenient to separate the expectation values of the wave amplitudes at frequency $\omega_0$ (denoted by uppercase letters) from their fluctuation components (denoted by lowercase letters), assuming that the fluctuations are small:
 $\hat c_0  \Rightarrow C_0 +  c_0 $.
Here, $C_0$ represents the expectation value of the field amplitude in the central optical mode, while $c_0$ denotes the quantum fluctuations of the field confined in this mode, with $|C_0|^2 \gg \langle c_0^\dag c_0 \rangle$. Similar expressions apply to the sideband optical modes as well as the mechanical mode. The amplitudes are normalized so that $\hslash \omega_0 |A_0|^2$ corresponds to the optical power of the pump.

We further assume that the expectation values of the amplitudes are real, as are the coupling constants $\eta_\pm$:
\begin{align}
	\label{real}
	A_0=A_0^*,\quad C_0= C_0^*=\sqrt{\frac{2}{\gamma_0}}A_0,\quad  \eta_\pm=\eta_\pm^*.
\end{align}

Since fluctuation waves around $\omega_0$ do not affect the field components near the frequencies $\omega_\pm$, and the first equation \eqref{c0} does not couple with the others. Therefore, we omit it from further consideration.

We use Fourier transform, introduce quadrature amplitudes for amplitude and phase
\begin{subequations}
\label{quadDef}
 \begin{align}
  a_{\pm a} = \frac{a_\pm (\Omega) +a_\pm ^\dag(-\Omega)}{\sqrt 2}\,,
	 a_{\pm \phi} = \frac{a_\pm (\Omega) -a_\pm ^\dag(-\Omega)}{i\sqrt 2}\,.
 \end{align}
\end{subequations}
and derive 
\begin{subequations}
	\label{quadAmp}
	\begin{align}
		\label{ca+DDNS}   
		(\gamma_+ - i\Omega)c_{+a} &+  \eta_+ C_0  d_a = \sqrt {2 \gamma_{0+}}  { a}_{ +a}  + \sqrt{2\gamma_{e+}} e_{+a},\\
		\label{ca-DDNS} 
		(\gamma_- - i\Omega) c_{-a} &- \eta_- C_0  d_a = \sqrt {2 \gamma_-}  { a}_{-a}+ \sqrt{2\gamma_{e-}} e_{-a},\\
		\label{daDDNS}
		(\gamma_m - i\Omega)  d_a &-  C_0 \Big(\eta_+c_{+a}+ \eta_-c_{-a}\Big)= \sqrt {2 \gamma_m} q_a +  f_{s\,a}.
	\end{align}
\end{subequations}
The set of equations \eqref{quadAmp} for the amplitude quadratures is independent of the set for the phase quadratures (not shown). 

Let us first consider the measurement of a narrowband ($\gamma_m \gg |\Omega| \rightarrow 0$) resonant force assuming that the thermal noise that comes with the classical force also represents the signal. This is the case which is physically realizable in electro-optical quantum transducers.

We are measuring the following linear combination of quadrature amplitudes 
\begin{align} \label{measquada}
\alpha_{a+} &= \frac{\sqrt{g_-} a_{+a} + \sqrt{g_+} a_{-a}}{\sqrt{g_++g_-}},\quad 
\alpha_{a-} =\frac{\sqrt{g_+} a_{+a} - \sqrt{g_-} a_{-a}}{\sqrt{g_++g_-}}, \\
        \label{measquadb}
    \beta_{a+} &= \frac{\sqrt{g_-} b_{+a} + \sqrt{g_+} b_{-a}}{\sqrt{g_++g_-}},\quad 
		\beta_{a-} =\frac{\sqrt{g_+} b_{+a} - \sqrt{g_-} b_{-a}}{\sqrt{g_++g_-}},
\end{align}
where the coefficients $g_\pm$ are identified in (\ref{zerodamp}); $b_{\pm a}$ are the amplitude quadratures for the field leaving the optical resonator, defined in the way identical with Eq.~(\ref{quadDef}) with replacement of $a_\pm\rightarrow b_\pm$. The linear combination of the amplitude quadratures of the output fields (\ref{measquadb}) is created as follows. We first measure both $b_{+a}$ and $b_{+a}$ quadratures separately and then evaluate arbitrary combinations of them via postprocessing of the measurement result. 

Analyzing the equations described above we arrive at the following relationships
\begin{align}
    \beta_{a+} &=\alpha_{a+}, \\
    \beta_{a-} &=\frac{\gamma_m -G }{\gamma_m+G}\alpha_{a-}
	-\frac{2(G^2+G_+^2)^{1/2}}{\gamma_m+G} \, \alpha_{a+} -\\
	 &\quad- \frac{\sqrt{2G_+}}{\gamma_m+G} \left(\sqrt {2 \gamma_m} q_a +  f_{s\,a}\right)
\end{align}
As we can see, the measurement result is not impacted by the optical noise if (i) we evaluate 
\begin{equation}
    \beta = \beta_{a-}+\frac{2(G^2+G_+^2)^{1/2}}{\gamma_m+G}\beta_{a+},
\end{equation}
and, (ii), select $ G=\gamma_m$.

\begin{figure}
\includegraphics[width=0.48\textwidth]{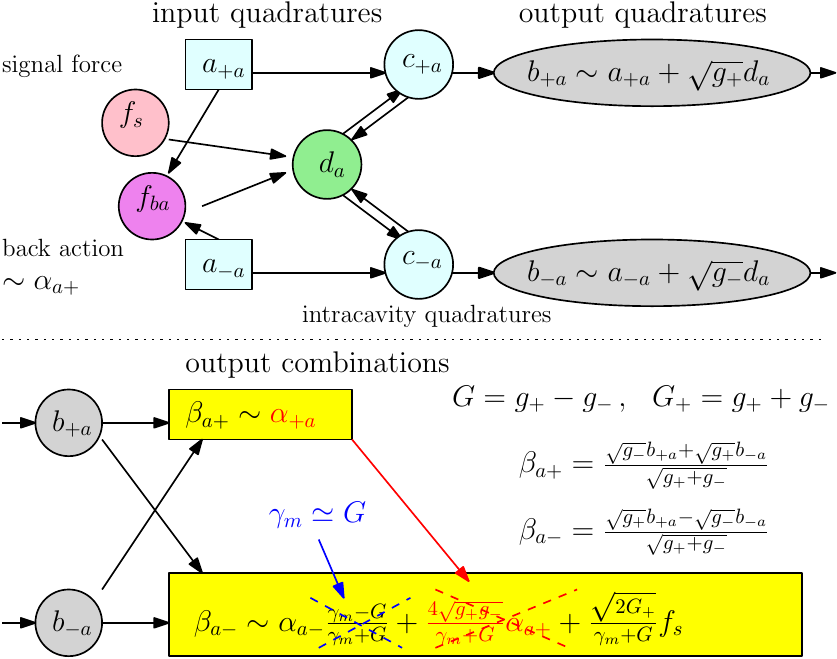}
	\caption{Illustration of measurement the output quadratures in the  asymmetric case: $\eta_\pm\ne\eta,\ \gamma_+\ne \gamma_-$.}\label{Quad_scheme2}
\end{figure}

In this case the measured value corresponds to the amplified combination of the signal force and the thermal noise associated with the force. 
\begin{equation}
    \beta=- \sqrt{\frac{G_+}{\gamma_m}} \left( q_a + \frac{ f_{s\,a}}{\sqrt {2 \gamma_m}}\right).
\end{equation}
This is the main observation of the current study. The procedure of the measurement is illustrated by Fig.~\ref{Quad_scheme2}.

There are two key considerations related to this result. First, in the electro-optical measurement scheme, the noise of the measured force can be attributed to the force itself and can be reduced if the electric signal is generated by a cold source. In contrast, in mechanical measurements, noise negatively impacts the measurement results. Second, the obtained result is strictly valid only for an infinitely long measurement ($\Omega \rightarrow 0$). To address these issues, we next analyze a more realistic scenario in which the force has a finite duration ($\gamma_m \ll \Omega$), which enables the elimination of thermal noise that overlaps with the signal at different spectral frequencies. 

In the simplest case of a lossless and symmetric system (same decays and the coupling rates) and $\Omega \geq \gamma_m$, the back action can be completely eliminated by optimally choosing the weight coeffcients of the separately measured and processed quadrature amplitudes. The spectral density of the quantum noise is
\begin{align}
	\label{symNoLoss}
	S_{qu}|_\text{noLoss}^\text{sym} &= \frac{\gamma_m^2 +\Omega^2}{\mathcal K}, \; \mathcal K=\frac{4\gamma\eta^2C_0^2}{\gamma^2+\Omega^2}. 
\end{align}
The spectral density monotonically decreases with power increase ($\sim \mathcal K$) and can be less than  $S_{SQL}=2(\gamma_m^2+\Omega^2)^{1/2}$.
\begin{figure}
\includegraphics[width=0.48\textwidth]{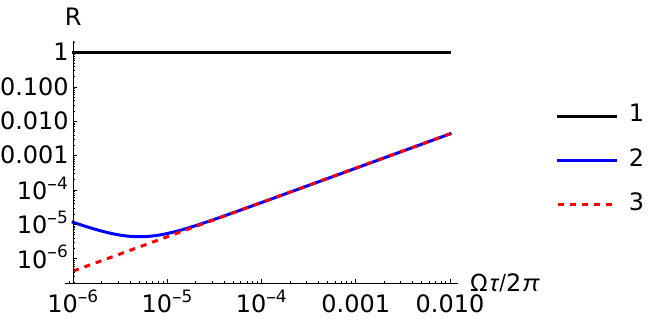}
	\caption{Plots of the ratio $R = S_{qu}/S_{SQL}$ of spectral densities, for the measurements without optical losses. Line (1) corresponds to  SQL, curve (2) --to  symmetric measurement scheme ($\gamma_+ = \gamma_-, \ \eta_+ = \eta_-$) and curve (3) -- to non-symmetric scheme with $G=\gamma_m$. Here $\tau$ is the duration of the signal force. }\label{NonSymNoLoss}
\end{figure}

For the non-symmetric case ($\eta_+\ne \eta_-$) and for the absence of optical losses ($\gamma_{e\pm}=0$) and nearly-resonant tuning, $\gamma_\pm\gg \Omega$ we find
\begin{align}
	\label{noLoss2}
	S_{qu}|_\text{noLoss}^\text{non-sym} &\underbrace{\simeq}_{\gamma_\pm \gg \Omega} \frac{1 }{2 G_+} \left(\big( \gamma_m - G)^2+\Omega^2 \right)
\end{align} 
The spectral density coincides with \eqref{symNoLoss} when $G=0$, meaning that in a lossless asymmetric measurement system, the back action can be completely eliminated. The normalized spectral density $S_{qu}$ decreases monotonically as the optical power increases.  Figure~(\ref{NonSymNoLoss}) illustrates the difference between ideal symmetrical measurement case and optimized non-symmetrical case. As expected, the introduced noise drops with $\Omega$ descrease.

\begin{figure}
\includegraphics[width=0.48\textwidth]{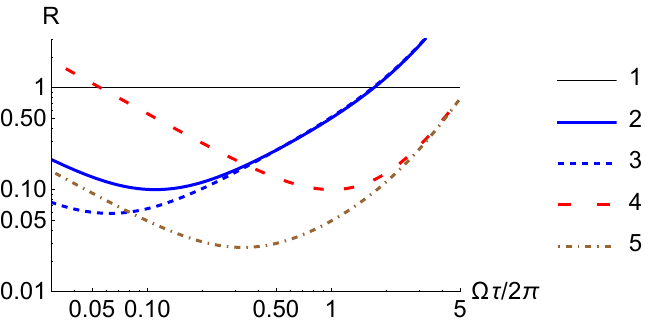}
	\caption{Plots of the ratio $R = S_{qu}/S_{SQL}$ of spectral densities for the system with small optical loss ($\gamma_e/\gamma_0=0.01$) for non-symmetric cases and  $\gamma_m\ll G$ as a function of spectral frequency $\Omega$ are shown. Line (1) stands for SQL, curve (2) corresponds to the measurement with residual optical loss (the dashed line (3) corresponds to the same parameters as curve (2) but with zero optical loss). Curves (4) and (5)  correspond to a 10 times larger input pump power as well as lossy and lossless cases, correspondingly. } \label{NonSymLoss2}
\end{figure}

For the condition $\gamma_m \ll G,\ \Omega$, we find
\begin{align}
	\label{noLoss3}
S_{qu}|_\text{noLoss}^\text{non-sym} &= \frac{G^2+\Omega^2}{2 G_+}\ge  \left|\frac{G\Omega }{ G_+} \right| < S_{SQL}\simeq 2|\Omega|.
\end{align}
Thus, although it is possible to surpass the SQL in the non-symmetric case, the sensitivity becomes limited. Also, as illustrated by Fig~(\ref{NonSymLoss2}), presence of the optical loss does not critically limit the measurement sensitivity. 

There are several practical challenges in the experimental realization of the proposed measurement scheme. These include: i) the difficulty of creating a triplet of optical modes with an equidistant frequency separation between $\omega_0$ and $\omega_\pm$; ii) achieving a frequency separation $|\omega_0 - \omega_\pm|$ equal to the mechanical frequency $\omega_m$ in the range of 1 to 100 MHz; iii) separating the optical harmonics without introducing coupling losses. This frequency difference ($\omega_m$) is quite small in the optical domain ($\omega_0$), and realizing such a narrow, lossless optical filter is a challenging experimental task. 
We can utilize optical modes from different geometric mode families along with corresponding geometric mode sorters \cite{Gu2018PRL, CarpenterNaureC2019} to implement the proposed system.

Consider higher-order modes (HOM) in a Fabry-Perot cavity. The frequency spacing between these modes can be made much smaller than the free spectral range (FSR) of the cavity by appropriately selecting the radii of curvature of the mirrors. The distribution functions of the optical modes, $\Psi_{00}(u)$, $\Psi_{01}(u)$, and $\Psi_{02}(u)$ can be optimized to enable the interaction. Here $u$ represents one of the transverse coordinates. The distributions over the other transverse coordinate $v$ are assumed to be Gaussian.

These optical modes can interact with the dipole elastic mode of the mirror (membrane). Distribution of the mechanical mode $\Psi_m$(u) should match the distribution of the optical modes. The generalized force acting on the elastic mode is determined by the overlap integral. In other words, the coupling constants $\eta_\pm$ are proportional to
\begin{subequations}
\label{etapm}
\begin{align}
 \eta_+ &\sim \int \Psi_{01}(u)\,\Psi_{02}(u)\, \Psi_m(u)\, du,\\
 \eta_- &\sim \int \Psi_{01}(u)\,\Psi_{00}(u)\, \Psi_m(u)\, du
\end{align}
\end{subequations}
The overlap integrals can be both nonzero and different, so that the absolute values of $\eta_+$ and $\eta_-$ also differ. Additionally, the increasing spatial width of the HOM leads to bigger diffraction losses. The damping rate changes from a mode to a mode. This asymmetry can enable the BAE method described above.

The next step is to spatially separate the modes leaving the FP cavity. The output light consists of radiation from all three optical modes. In theory, a mode sorter can achieve perfect separation, and experimental efficiencies exceeding 80\% have recently been confirmed \cite{Gu2018PRL}. Therefore, the modes can be separated, reshaped into Gaussian beams, and optimally detected to implement the proposed measurement.

In conclusion, we have explored the optimization of quantum measurements for a classical force resonant with a mechanical mode in an optomechanical transducer. This transducer consists of an optical Fabry-Perot resonator, where one of the mirrors is suspended on a resonant mechanical structure. The applied force generates optical harmonics shifted from the optical pump frequency by the force’s frequency. Our analysis reveals that by examining both optical harmonics, it is possible to perform back-action-evading measurements of the force having better sensitivity than traditional back action evading measurement. The optimal performance is achieved when the cavity modes exhibit different optical damping rates and distinct optimal overlap integrals with the mechanical mode. This asymmetry in optical couplings introduces an additional damping term, which plays a crucial role in suppressing quantum back-action. 

\acknowledgements

 The research of AAM, AIN and SPV has been supported by  Theoretical Physics and Mathematics Advancement Foundation “BASIS” (Contract No. 22-1-1-47-1), by the Interdisciplinary Scientific and Educational School of Moscow University ``Fundamental and Applied Space Research'' and by the TAPIR GIFT MSU Support of the California Institute of Technology. The reported here research performed by ABM was carried out at the Jet Propulsion Laboratory, California Institute of Technology, under a contract with the National Aeronautics and Space Administration (80NM0018D0004).  This document has LIGO number P2400323.

%


\begin{thebibliography}{99}

\bibitem{Braginsky68}
V. B. Braginsky, Classic and quantum limits for detection of weak force on acting on macroscopic oscillator, Sov. Phys. JETP {\bf 26}, 831 (1968).

\bibitem{92BookBrKh}
V. B. Braginsky and F. Y. Khalili, {\em Quantum measurement,} (Cambridge University Press, 1995).

\bibitem{Povinelli05ol}
M. L. Povinelli, M. Loncar, M. Ibanescu, E. J.  Smythe, S. G. Johnson, F. Capasso,  and J. D. Joannopoulos, Evanescent-wave bonding between optical waveguides, Opt. Lett. {\bf 30}, 3042 (2005).

\bibitem{maslov13pra}
A. V. Maslov, V. N. Astratov, and M. I. Bakunov, Resonant propulsion of a microparticle by a surface wave, Phys. Rev.A {\bf 87}, 053848 (2013).

\bibitem{21a1VyNaMaPRA}
S. P. Vyatchanin, A. I. Nazmiev, and A. B. Matsko, Broadband dichromatic variational measurement, Phys. Rev. A {\bf 104}, 023519 (2021).

\bibitem{Gu2018PRL}
X. Gu, M. Krenn, M. Erhard, and A. Zeilinger, Gouy phase radial mode sorter for light: concepts and experiments, Phys. Rev. Lett. {\bf 120}, 103601 (2018).

\bibitem{CarpenterNaureC2019}
N. K. Fontaine, R. Ryf, H. Chen,  D. T. Neilson, K. Kim,  and J. Carpenter, J., Laguerre-Gaussian mode sorter, Nature Comm. {\bf 10}, 1865 (2019).

\end{thebibliography}

\end{document}